\documentclass[sigconf]{acmart}

\setcopyright{none}
\renewcommand\footnotetextcopyrightpermission[1]{}
\acmYear{2026}
\acmDOI{}
\acmISBN{}

\usepackage{booktabs}   
\usepackage{graphicx}   
\usepackage{amsmath}    
\usepackage{algorithm}
\usepackage{algpseudocode}
\usepackage{microtype}
\begin{document}
\pagestyle{plain} 

\title{From Sessions to Serving: Integrating LLM-Based Embeddings into Retrieval for Production Content Recommendation}


\author{Zhixiong Hu}
\affiliation{%
  \institution{Autodesk}
  \city{San Francisco}
  \state{CA}
  \country{USA}
}

\author{Meng Li}
\affiliation{%
  \institution{Autodesk}
  \city{San Francisco}
  \state{CA}
  \country{USA}
}

\author{Maggie Zhang}
\affiliation{%
  \institution{Autodesk}
  \city{San Francisco}
  \state{CA}
  \country{USA}
}

\author{Deborshi Goswami}
\affiliation{%
  \institution{Autodesk}
  \city{San Francisco}
  \state{CA}
  \country{USA}
}

\author{David Bailey}
\affiliation{%
  \institution{Autodesk}
  \city{San Francisco}
  \state{CA}
  \country{USA}
}

\author{James Bradley}
\affiliation{%
  \institution{Autodesk}
  \city{San Francisco}
  \state{CA}
  \country{USA}
}

\begin{abstract}
Personalized content recommendation on large marketplace websites must
reconcile heterogeneous candidate content (e.g., customer stories,
product pages, industry pages, and learning content) with sparse,
fast-changing user behavior, placing increasing demands on the
\emph{retrieval} stage to represent user interest well, not just the
downstream ranking stage. We present LSER (LLM Session Embedding
Retrieval), a production content recommendation system for
autodesk.com that derives session- and history-level embeddings from a
pretrained large language model (LLM) and \emph{adds} them as auxiliary
user-side features in the retrieval-stage user tower alongside existing learned sequence encoders. To keep user-side embeddings fresh under real-time behavior, we build a
nearline serving path that asynchronously refreshes sequential and
LLM-derived user vectors for our main content recommendation server. We validate the incremental value
of the LLM features through an ablation study across our full
multi-content-type catalog.
Our results indicate that integrating LLM-derived embeddings as
additional retrieval signals, together with production-grade nearline
serving, is technically feasible and can improve model performances.
\end{abstract}




\maketitle

\section{Introduction}
\label{sec:intro}

Personalized content recommendation on large marketplace websites such as
autodesk.com must contend with highly heterogeneous candidate pools
(customer stories, product pages, industry pages, and learning content),
sparse and bursty real-time behavioral signals, and the need to reconcile
short-term session intent with long-term user context (e.g., product
usage history). Classical two-stage retrieval-and-ranking pipelines
address scale by first retrieving a high-recall candidate set and then
ranking it with a learned model~\cite{huang2020embedding}, but the
quality of both stages is bottlenecked by how well the \emph{retrieval}
stage represents a user's evolving interests, not just the ranking
stage.

Recent work has shown that embeddings derived from large language models
(LLMs) carry rich semantic signal that complements traditional ID-based
collaborative representations in sequential and session-based
recommendation. Boz et al.~\cite{boz2025improving}
demonstrate several ways of injecting LLM-derived item embeddings into
sequential recommenders, including aggregating them into a session-level
embedding for direct similarity-based retrieval. Liu et
al.~\cite{liu2024llmesr} use frozen LLM semantic embeddings together with
retrieval-augmented self-distillation to address long-tail user and item
coverage in sequential recommendation, and Cui et
al.~\cite{cui2025semantic} similarly use LLM-derived semantic embeddings
to retrieve informative contrastive pairs for sequence encoder training.
Kim et al.~\cite{kim2024allmrec} instead let an LLM directly consume the
user/item embeddings of a pretrained collaborative filtering model,
avoiding the extensive fine-tuning that LLM-based recommenders typically
require. On the industrial side, LEARN~\cite{jia2025learn} freezes a
pretrained LLM as an item encoder and aligns it to a twin-tower
recommendation objective at industrial scale on a short-video
platform, while
TreeBridge~\cite{ni2026treebridge} bridges the semantic/behavioral gap
between LLM embeddings and collaborative signals at Shopee with an
online-offline hybrid serving paradigm, reporting a measurable lift in
gross merchandise volume after deployment.

Beyond embedding-level integration, a parallel line of industrial work
has pushed retrieval and ranking architectures themselves toward
greater scale and expressiveness. Zhang et al.~\cite{zhang2024wukong}
establish a scaling law for large-scale recommendation models
analogous to those observed for LLMs, and Zhai et
al.~\cite{zhai2024actions} reformulate recommendation as a generative
sequential transduction task with their HSTU architecture, scaling to
trillions of parameters and demonstrating a power-law relationship
between model quality and training compute, while
OneRec~\cite{deng2025onerec} replaces the conventional cascaded
retrieve-then-rank pipeline with a single end-to-end generative model.
PinRec~\cite{botta2025pinrec} and GRank~\cite{sun2026grank} likewise
adopt generative retrieval at industrial scale, the latter unifying
target-aware candidate generation with lightweight ranking to avoid the
maintenance cost of structured indices. Complementary to generative
retrieval, Zhang et al.~\cite{zhang2025mtmh} show that jointly optimizing
recall and semantic relevance with a multi-head item-to-item retrieval
architecture meaningfully improves both engagement and content
diversity, Fan et al.~\cite{fan2025pinterest} propose a multi-embedding
retrieval framework that represents a single user with multiple
embeddings conditioned on implicit and explicit interests, and Lyu et
al.~\cite{lyu2025dv365} show that incorporating an offline-computed,
extremely long user history embedding is highly incremental on top of
existing attentive sequence models deployed across multiple production
surfaces. Together, these systems illustrate that the retrieval
stage itself — not only ranking — is an increasingly active area where
richer user/session representations, including those derived from
LLMs, are being deployed at industrial scale.

These results collectively confirm that LLM-derived session and user
representations, and richer retrieval architectures more broadly, can
meaningfully improve production recommender
systems~\cite{jia2025learn,ni2026treebridge,zhai2024actions,deng2025onerec,zhang2025mtmh,fan2025pinterest,lyu2025dv365}.
However, each of these systems was developed and
tuned for a specific company's data, content catalog, and serving
infrastructure, and the concrete design decisions (how the session
encoder is adapted and fine-tuned, how its output is integrated into
an existing retrieval stage, and how the resulting embedding is served
with production latency and freshness constraints) are necessarily
specific to each deployment context. We see real value in reporting a
concrete, end-to-end account of how this integration was carried out
in our own setting, to add to the body of practical, reproducible
evidence available to the community.

In this paper, we present LSER, a system built around our own
multi-content-type recommendation setting on autodesk.com and the
constraints of our existing serving infrastructure. We adapt a pretrained LLM to produce session- and history-level
embedding vectors, concatenate them into our retrieval-stage user
tower as \emph{additional} features (while retaining learned
in-tower sequence encoders for production), and build a nearline
serving path that refreshes learned sequential and LLM-derived user
vectors in time for our main content recommendation server.
We validate their incremental value through an offline ablation across
our full multi-content-type catalog, and further observe encouraging
improvements in engagement and downstream conversion metrics in an
online A/B test across several production surfaces.
Our main contributions are:
\begin{enumerate}
  \item We design an LLM-based embedding path that produces auxiliary
  session and history vectors, adapt a pretrained LLM with custom
  modifications and a fine-tuning strategy tailored to our
  multi-content-type setting, and integrate these vectors into the
  retrieval user tower without replacing existing sequence encoders.
  \item We deploy nearline refresh of user-side embeddings for production
  serving and validate the system with offline and online experiments.
\end{enumerate}

\section{Method}
\label{sec:method}

\subsection{Retrieval Architecture}
\label{sec:method-retrieval}

\begin{figure*}[t]
  \centering
  \includegraphics[width=1\textwidth]{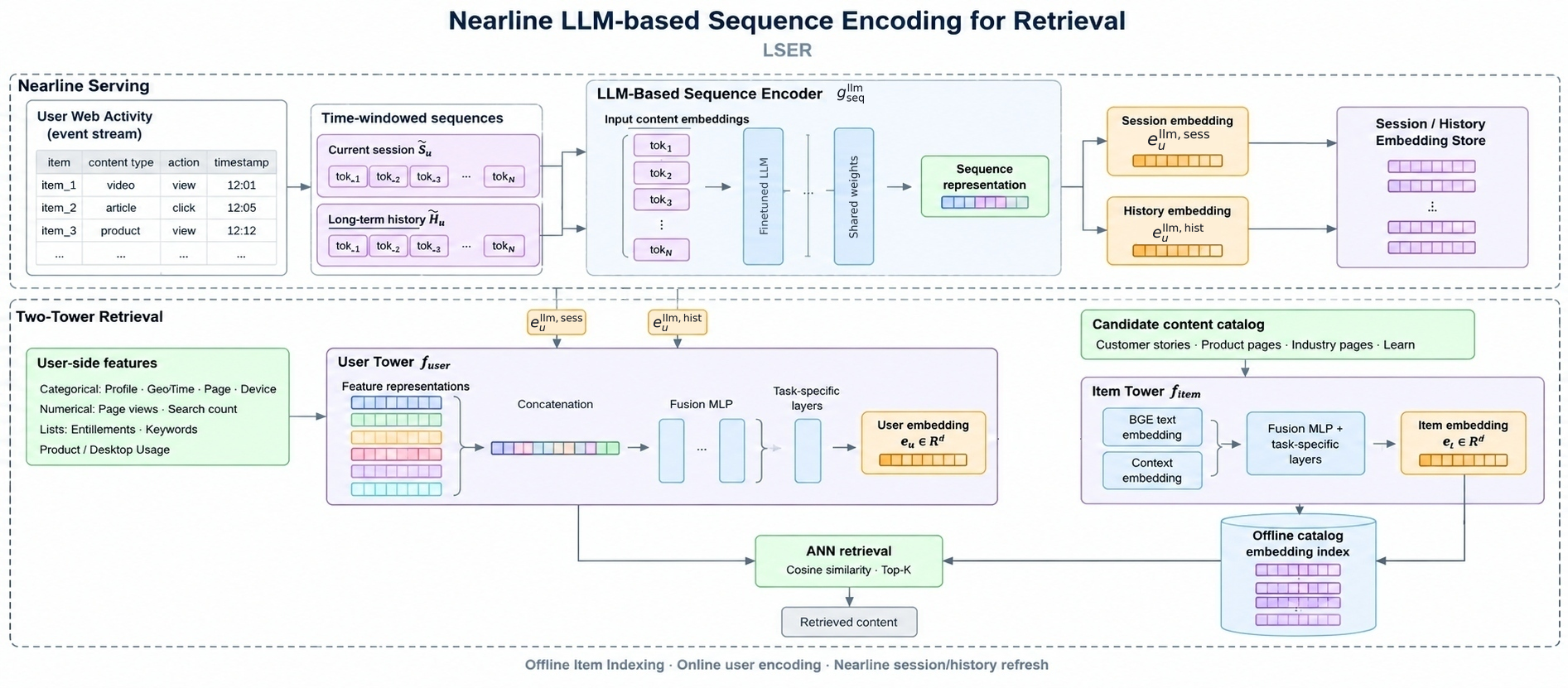}
  \caption{Overview of the LSER architecture. A nearline serving
  pipeline (top) asynchronously refreshes user-side embedding components
  from recent activity and writes them to an embedding store.
  At serving time, a two-tower retrieval model (bottom) assembles a
  user embedding $\mathbf{e}_u$ in the User Tower $f_{\text{user}}$ and
  an item embedding $\mathbf{e}_i$ in the Item Tower $f_{\text{item}}$;
  retrieval ranks candidates by cosine similarity
  (Section~\ref{sec:method-retrieval}) for Top-$K$ results.}
  \Description{System diagram with a nearline path that refreshes user
  embeddings from session and history activity into a store, and a
  two-tower retrieval path that fuses stored vectors with request-time
  features to rank items by similarity.}
  \label{fig:architecture}
\end{figure*}

LSER performs retrieval with a two-tower architecture: a \emph{user
tower} that encodes a user's short- and long-term interests into a
single dense representation, and an \emph{item tower} that encodes
each candidate piece of content into a representation of the same
dimensionality. Retrieval reduces to an approximate nearest-neighbor
search between the user vector and the item vectors in the catalog,
which scales to our full multi-content-type candidate pool (customer
stories, product pages, industry pages, and learning content) at
production latency. This retrieval stage produces a smaller candidate
set of Top-$K$ items, which is then passed to a downstream ranking
step (outside the scope of this paper) that applies additional,
more fine-grained scoring before final content is shown to the user.
Figure~\ref{fig:architecture} gives an overview of the full system,
spanning both the nearline serving pipeline and the two-tower retrieval
model described in this section.

\subsubsection{Notation}
Let $u \in \mathcal{U}$ denote a user and $i \in \mathcal{I}$ denote a
candidate content item, where $\mathcal{I}$ spans all supported
content types. At a label or request time $t_0$, user $u$ is described
by sequential and non-sequential features. Let
$S_u = (s_1, \ldots, s_{n_s})$ and $H_u = (h_1, \ldots, h_{n_h})$ denote the
session and history \emph{item-id sequences} consumed by the learned
sequence encoder $g_{\text{seq}}$, where $n_s$ and $n_h$ are the session
and history sequence lengths (capped in practice). For
$k \in \{1,\ldots,n_s\}$, $s_k$ is the catalog item identifier at
chronological position $k$ in $S_u$; for $m \in \{1,\ldots,n_h\}$, $h_m$
is defined analogously at position $m$ in $H_u$. Each entry is a catalog item id or an \emph{unknown} placeholder when mapping
fails. The lists are constructed point-in-time from User Web Activity as
in Section~\ref{sec:method-sequences}. Intuitively, $S_u$ covers in-session
behavior before $t_0$; $H_u$ covers earlier activity before the current
session starts. The optional LLM auxiliary path
(Section~\ref{sec:method-encoder}) uses the \emph{same} time windows but
serializes \emph{raw} logged event fields rather than these item-id lists. Non-sequential user features include categorical
$x_u^{\text{cat}}$, numerical $x_u^{\text{num}}$, list-valued
$x_u^{\text{list}}$, and product/desktop usage signals (collectively
denoted $x_u$), as in Section~\ref{sec:method-data}. On the item side, $x_i^{\text{text}}$ and
$x_i^{\text{context}}$ denote textual content and contextual metadata.
We use $i^+$ (or $i_k^+$ in batch indices) for the supervised positive
item in a label receipt. Learned embedding modules are
$\mathrm{Emb}(\cdot)$; fusion/MLP towers are $f_{\text{item}}$,
$f_{\text{user}}$; the sequence encoder is $g_{\text{seq}}$; user and
item embeddings are $\mathbf{e}_u, \mathbf{e}_i \in \mathbb{R}^d$ with
session/history components $\mathbf{e}_u^{\text{sess}}$ and
$\mathbf{e}_u^{\text{hist}}$. Vector concatenation is
$[\cdot \, ; \, \cdot]$.

\subsubsection{Data and Feature Construction}
\label{sec:method-data}
We represent each raw user interaction as an event tuple
\begin{equation}
  e = (\,\mathrm{item\_id}, \; \mathrm{content\_type}, \;
  \mathrm{action\_type}, \; t\,),
  \label{eq:event}
\end{equation}
where $\mathrm{content\_type}$ ranges over the full set of content
types available on autodesk.com (e.g., products, industries, customer
stories, and learn content), $\mathrm{action\_type}$ distinguishes
interaction strength (e.g., impression, click, trial/install, order),
and $t$ is the event timestamp. In production, \emph{User Web Activity}
stores richer records (e.g., page URL, title, locale, page category,
device type, optional content identifier, and click/link metadata);
Eq.~\eqref{eq:event} is the catalog-normalized view used once an event
is mapped to a content item. These logs are the primary source from
which session and history features (Section~\ref{sec:method-sequences})
are derived.

\paragraph{Constructing session and history sequences.}
\label{sec:method-sequences}
Sequential features are built \emph{point-in-time} for each supervised
row or live request. We write $t_0$ for that row's label-receipt
timestamp (offline) or the recommendation request time (online). All
events with timestamp $t \ge t_0$ are excluded so that sequences never
leak future behavior relative to the label.

Events are drawn from User Web Activity using the point-in-time rules
below. For $g_{\text{seq}}$, we apply channel-specific eligibility
filters (below), map each retained event to a catalog item identifier,
and form chronological item-id lists; events without a resolvable catalog
item are dropped or mapped to a generic unknown token. We use the logged
session identifier to obtain the current session start
$t_{\mathrm{start}}$ (the earliest event time in that session).

\textbf{Session sequence ($S_u$).} Collect eligible engagement events
for the same user and session with
$t \in [t_{\mathrm{start}}, t_0)$, sort by $t$, and retain the most
recent events up to a fixed maximum length. Intuitively, $S_u$ summarizes
in-session behavior immediately before the labeled action or request.

\textbf{History sequence ($H_u$).} Collect eligible events for the same
user from earlier activity: timestamps in a configurable multi-day
lookback window ending at $t_{\mathrm{start}}$, sorted and truncated
with a separate (typically longer) length cap. This separates
short-horizon in-session intent from longer-horizon return behavior.

\textbf{Click and learn channels.} Marketplace and learning-content
signals are materialized as parallel session and history sequence
fields (four item-id lists in implementation) that share the
time-slicing rules above but apply channel-specific eligibility
filters before catalog items enter a list. The user tower encodes each
channel and fuses the resulting vectors with non-sequential features
(Section~\ref{sec:method-retrieval}). We use $S_u$ and $H_u$ for the session-side and history-side item-id
lists; Eq.~\ref{eq:seq-encoder} applies $g_{\text{seq}}$ after mapping
each list to content embedding sequences $\mathbf{X}_u^{\text{sess}}$ and
$\mathbf{X}_u^{\text{hist}}$ (and any within-tower channel fusion). Richer event serialization for the LLM
auxiliary encoder over the analogous windows is described in
Section~\ref{sec:method-encoder}.

\textbf{Offline event coverage.} When building the training table, we
materialize label receipts over the experiment date range and ingest
raw events over a sufficiently long pre-label window so every row can
reconstruct $(S_u, H_u)$ point-in-time without using post-$t_0$ events.
Train/validation/test splits are applied on receipt time
(Section~\ref{sec:experiments-offline}).

At serving time the same rules apply with $t_0$ set to request time;
session and history lists are refreshed nearline
(Section~\ref{sec:serving}). This mirrors the pattern of deriving
short- and long-horizon views from one timeline, as in long-history
recommender feature stacks~\cite{lyu2025dv365}, but with session-aware
boundaries rather than a single fixed short window anchored only at
$t_0$.

Beyond these sequential features, we construct three further
non-sequential user-side feature families:

\begin{itemize}
  \item \textbf{Categorical features} $x_u^{\text{cat}}$: static or
  slowly-changing single-valued attributes, including User Profile
  (account-level attributes), Geo \& Time (user location and
  request-time context), Page Context (the surface/page a request
  originates from), and Device Type.
  \item \textbf{Numerical features} $x_u^{\text{num}}$: scalar
  engagement counters such as recent page-view count and search count.
  \item \textbf{List features} $x_u^{\text{list}}$: multi-valued,
  variable-length attributes, including Entitlements (the set of
  Autodesk products and licenses a user already owns or has access to,
  which is important for distinguishing content that is novel/relevant
  to a user from content about products they already use) as well as
  page-view and search keywords (free-text signals capturing on-site
  search and browsing intent).
\end{itemize}

In addition to the above, we incorporate a Product/Desktop Usage
feature family capturing longer-horizon, lower-frequency
signals about how a user engages with the Autodesk desktop products
they have access to (e.g., which product areas and, where available,
which commands or features within a product a user has recently used).
These product usage signals complement the primarily web-activity-driven
feature families above with a view into in-product behavior, and are
folded into the user tower's non-sequential feature inputs alongside
$x_u^{\text{cat}}$, $x_u^{\text{num}}$, and $x_u^{\text{list}}$.

On the item side, we construct a textual content feature
$x_i^{\text{text}}$ (the item's title and description text) and a
context feature $x_i^{\text{context}}$ (categorical metadata such as
content type). These are consumed by the item tower described next;
since our main architectural contribution concerns the \emph{user}
side of retrieval, we describe the item tower only briefly before
turning to the user tower in detail.

\subsubsection{Item Tower}
The item tower produces an embedding $\mathbf{e}_i \in \mathbb{R}^d$
for each candidate item $i$ by encoding its text and context features
and fusing them together:
\begin{equation}
  \mathbf{e}_i = f_{\text{item}}\big(
    \mathrm{Emb}_{\text{text}}(x_i^{\text{text}}), \;
    \mathrm{Emb}_{\text{context}}(x_i^{\text{context}})
  \big),
  \label{eq:item-tower}
\end{equation}
where $\mathrm{Emb}_{\text{text}}(\cdot)$ encodes title/description
text using a pretrained BGE (BAAI General
Embedding)~\cite{xiao2023cpack} text encoder, $\mathrm{Emb}_{\text{context}}(\cdot)$
embeds categorical context features, and $f_{\text{item}}$ is a fusion
network (an MLP followed by task-specific transformation layers) that
combines the two into a single item embedding $\mathbf{e}_i$. Our
architecture also supports an optional, learned item-ID embedding
alongside the above, but we found it unnecessary given sufficiently
rich text-based item representations, and it is disabled in
production; $\mathbf{e}_i$ is therefore derived purely from content
rather than from a memorized item identity, which we expect to also
aid generalization to newly onboarded items. We note that this
item-side design is a relatively standard content embedding setup; it
is the user tower below that houses our main architectural
contribution.

\subsubsection{User Tower}
The user tower is where our main architectural contribution lies: it
is responsible for synthesizing short-term session intent, long-term
history, and auxiliary user-side signals into a single representation
that is used to retrieve relevant content, and it is the component
into which LLM-derived auxiliary embeddings (Section
\ref{sec:method-encoder}) are integrated. The user tower combines
signal from the sequential and non-sequential feature families
introduced in Section~\ref{sec:method-data}: (i) a \emph{session
embedding} summarizing the user's current, real-time intent, (ii) a
\emph{history embedding} summarizing longer-term interests, (iii)
optional \emph{LLM-derived} session and history vectors served
nearline (Section~\ref{sec:method-encoder}), and (iv) the
non-sequential categorical, numerical, list, and product/desktop
usage feature families. Session and history are item-id sequences
$S_u$ and $H_u$ (Section~\ref{sec:method-sequences}). Before
$g_{\text{seq}}$, each position is mapped to a \emph{content
embedding} vector: for $s_k \in S_u$ (and analogously $h_m \in H_u$),
we look up the catalog item's title/description text embedding from
the same pretrained BGE encoder that defines
$\mathrm{Emb}_{\text{text}}(\cdot)$ in Eq.~\ref{eq:item-tower}, and
stack the vectors in time order with left padding to a fixed length,
producing padded sequence tensors $\mathbf{X}_u^{\text{sess}}$ and
$\mathbf{X}_u^{\text{hist}}$ plus boolean validity masks (padding
positions are masked out). A learned sequence encoder
$g_{\text{seq}}$ (BERT4Rec-style transformer over embedded sequences
in our initial production deployment) maps each tensor to a fixed-length
vector via masked pooling over its outputs:
\begin{equation}
  \mathbf{e}_u^{\text{sess}} = g_{\text{seq}}(\mathbf{X}_u^{\text{sess}}),
  \qquad
  \mathbf{e}_u^{\text{hist}} = g_{\text{seq}}(\mathbf{X}_u^{\text{hist}}),
  \label{eq:seq-encoder}
\end{equation}
We optionally concatenate a learned item-id embedding at each position
when enabled; in production we rely on the shared BGE content vectors
alone (consistent with disabling memorized item-id embeddings in
Eq.~\ref{eq:item-tower}).
Let $\mathbf{e}_u^{\text{LLM}} = [\mathbf{e}_u^{\text{LLM,sess}};
\mathbf{e}_u^{\text{LLM,hist}}]$ denote the concatenation of
nearline LLM-derived session and history embeddings when enabled
(Section~\ref{sec:method-encoder}, Eq.~\ref{eq:llm-encoder}). In production, the sequential and
LLM-derived components of the user tower are computed \emph{nearline}
(Section~\ref{sec:serving}) and read at request time; training still
runs $f_{\text{user}}$ end-to-end on the same tensors. The non-sequential feature
families are each encoded with an operator suited to their structure:
categorical features via embedding lookup, numerical features via an
MLP, and list features (including Entitlements and the
page-view/search keyword lists, as well as the product/desktop usage
feature family) via pooled or attention-based set encoders such as an
embedding bag or multi-head attention over the list elements. Let
$\mathrm{Emb}_{\text{other}}(\cdot)$ denote the encoded bundle of all
non-sequential user features in Section~\ref{sec:method-data}
(categorical, numerical, list, and product/desktop usage). The sequence
embeddings, optional $\mathbf{e}_u^{\text{LLM}}$, and
$\mathrm{Emb}_{\text{other}}$ are fused through $f_{\text{user}}$ (an
MLP followed by task-specific transformation layers) to produce the
final user embedding:
\begin{equation}
  \mathbf{e}_u = f_{\text{user}}\big(
    \mathbf{e}_u^{\text{sess}}, \;
    \mathbf{e}_u^{\text{hist}}, \;
    \mathbf{e}_u^{\text{LLM}}, \;
    \mathrm{Emb}_{\text{other}}(x_u)
  \big) \in \mathbb{R}^d.
  \label{eq:user-tower}
\end{equation}
When LLM features are disabled, $\mathbf{e}_u^{\text{LLM}}$ is omitted
and $f_{\text{user}}$ matches the production baseline configuration.
Because $\mathbf{e}_u$ and $\mathbf{e}_i$ are produced by separate
towers that only interact through a similarity function at the output
(rather than through early cross-attention between user and item
features), item embeddings are precomputed and indexed offline, while
the expensive user-side encoders run nearline and the request path
performs lightweight fusion in $f_{\text{user}}$ plus ANN scoring---which
keeps large-scale retrieval within latency budgets
(Section~\ref{sec:serving}).

\subsubsection{Scoring and Training Objective}
Given a user embedding $\mathbf{e}_u$ and an item embedding
$\mathbf{e}_i$, we score their compatibility with cosine similarity,
$\mathrm{score}(u, i) = \mathbf{e}_u^\top \mathbf{e}_i /
(\lVert \mathbf{e}_u \rVert \, \lVert \mathbf{e}_i \rVert)$.

\paragraph{Positive supervision via label receipts.}
\label{sec:method-labels}
Training examples are supervised $(u, i^+)$ pairs whose positive item
is not defined from a single raw log field in isolation. Each instance
is associated with a \emph{label receipt}: a consolidated supervision
record that aggregates outcome and context information from multiple
internal telemetry and labeling sources (e.g., on-site behavioral
events, content and surface metadata, and other eligible signals where
available). Lightweight fusion and eligibility rules over these
receipts map each record to one canonical item identity and content
type, together with the user-side feature snapshot at label time.
This lets heterogeneous engagement definitions across content types and
surfaces contribute to the same contrastive retrieval objective without
maintaining separate training corpora or losses per event family. We
do not expose receipt schemas here; representative positive families
include strong web-engagement outcomes on marketplace and learning
surfaces, among others admitted by the fusion rules.

We train both towers jointly end-to-end with an InfoNCE
contrastive loss~\cite{huang2020embedding} using in-batch negatives.
For a training batch of $N$ (user, positive item) pairs
$\{(u_k, i_k^+)\}_{k=1}^{N}$, the loss for example $k$ is
\begin{equation}
  \mathcal{L}_k = -\log \frac{
    \exp\big(\mathrm{score}(u_k, i_k^+) / \tau\big)
  }{
    \sum_{j=1}^{N} \exp\big(\mathrm{score}(u_k, i_j^+) / \tau\big)
  },
  \label{eq:infonce}
\end{equation}
where $i_j^+$ is the positive item for batch index $j$ (so for
$j \neq k$, $i_j^+$ is an in-batch negative for user $u_k$), and
$\tau$ is a learned or fixed temperature
hyperparameter controlling the sharpness of the softmax. The overall
training objective is the batch-averaged loss
$\mathcal{L} = \frac{1}{N}\sum_{k=1}^{N} \mathcal{L}_k$.

At serving time, the trained item tower is applied once, offline, to
the full item catalog to produce an index of item embeddings, while
the user tower consumes nearline-refreshed session, history, and LLM
vectors together with request-time features (Section~\ref{sec:serving});
retrieval returns the top-$K$ items by
$\mathrm{score}(u, i)$ via approximate nearest-neighbor search over
this index.

Our main modeling contribution for LLMs is not to replace
$g_{\text{seq}}$, but to add $g_{\text{seq}}^{\text{LLM}}$ and
$\mathbf{e}_u^{\text{LLM}}$ in Eqs.~\ref{eq:llm-encoder}
and~\ref{eq:user-tower}: a complementary semantic view of user behavior
over the same session/history windows as Section~\ref{sec:method-sequences},
produced offline/nearline and consumed like other user-side features at
retrieval time. Section~\ref{sec:experiments-offline}
ablations hold $g_{\text{seq}}$ and the rest of the tower fixed and
toggle only these LLM-derived vectors.

\subsection{LLM-Derived Session and History Embeddings}
\label{sec:method-encoder}

This subsection defines the LLM-based sequence encoder
$g_{\text{seq}}^{\text{LLM}}$ (parallel to $g_{\text{seq}}$ in
Eq.~\ref{eq:seq-encoder}) and produces $\mathbf{e}_u^{\text{LLM,sess}}$,
$\mathbf{e}_u^{\text{LLM,hist}}$, and $\mathbf{e}_u^{\text{LLM}}$ in
Eq.~\ref{eq:llm-encoder}. At a high level, raw User Web Activity events from the same point-in-time
session and history windows as Section~\ref{sec:method-sequences}---not
the item-id sequences fed to $g_{\text{seq}}$---are converted into
event-token sequences for a pretrained LLM backbone; the backbone is
adapted (rather than trained from scratch) for autodesk.com's
multi-content-type logs, fine-tuned with a session-focused objective,
and pooled into fixed-length vectors that are \emph{concatenated} into
$f_{\text{user}}$ alongside $\mathbf{e}_u^{\text{sess}}$ and
$\mathbf{e}_u^{\text{hist}}$ from $g_{\text{seq}}$. Nearline serving of
$\mathbf{e}_u^{\text{LLM}}$ is described in Section~\ref{sec:serving},
not here.

\subsubsection{Input Representation}
\label{sec:method-llm-input}
Let $\widetilde{S}_u$ and $\widetilde{H}_u$ denote the session- and
history-side \emph{event-token sequences} for the LLM path (the same
point-in-time windows as $S_u$ and $H_u$ in
Section~\ref{sec:method-sequences}, but not the item-id lists consumed
by $g_{\text{seq}}$). Each position is a web event encoded as
(\emph{event type}, \emph{event value}) text: event type is the user
action (e.g., page view, link click) and event value describes the page
or content context. Events are ordered by timestamp (ascending). We
append a special end-of-sequence token \texttt{<EOS>}, pad or truncate
to a fixed length, and build an attention mask that excludes padded
positions. Event type and event value text at each position are mapped to vectors
with a frozen pretrained text embedding model (Qwen3-0.6B-embedding in
our deployment), yielding $e_k^{\text{type}}$ and $e_k^{\text{val}}$;
their concatenation $\mathbf{z}_k = [e_k^{\text{type}}; e_k^{\text{val}}]$
is the continuous input vector at timestep $k$. Stacking
$(\mathbf{z}_1,\ldots,\mathbf{z}_{n_s})$ in time order with the same
padding and attention mask yields the session-side matrix
$\widetilde{\mathbf{X}}_u^{\text{sess}}$ (and
$\widetilde{\mathbf{X}}_u^{\text{hist}}$ analogously from
$\widetilde{H}_u$).

\subsubsection{Backbone and Adaptation}
The backbone is a Qwen3-style~\cite{qwen2025qwen3} decoder stack with 16 layers and 4
attention heads per layer. Event type and value are already embedded into
continuous vectors $\mathbf{z}_k$ in Section~\ref{sec:method-llm-input}, so
the stack consumes $\widetilde{\mathbf{X}}_u^{\text{sess}}$ and
$\widetilde{\mathbf{X}}_u^{\text{hist}}$ rather than discrete token ids.
We therefore replace the vocabulary embedding lookup with a linear map
from the per-timestep event embedding dimension (size of $\mathbf{z}_k$)
to hidden size. The model is designed to consume event embeddings rather than event ids for scalability, considering growing events over time.
Rotary position embedding (RoPE)~\cite{su2024roformer} encodes order
within $\widetilde{\mathbf{X}}_u^{\text{sess}}$ and
$\widetilde{\mathbf{X}}_u^{\text{hist}}$.

\subsubsection{Training Objective and Data Construction}
We fine-tune the LLM sequence encoders $g_{\text{seq}}^{\text{LLM}}$ with a
\emph{next-event prediction} objective (session and history copies in
Eq.~\ref{eq:llm-encoder} are trained with the same recipe on their
respective windows). Each training instance is built from logged web
events by choosing a context prefix and the event that occurs immediately
afterward. Instances are sampled from either the session-side matrix
$\widetilde{\mathbf{X}}_u^{\text{sess}}$ or the history-side matrix
$\widetilde{\mathbf{X}}_u^{\text{hist}}$ in
Section~\ref{sec:method-llm-input}; \emph{history is handled the same
way as session}---only the source sequence differs. Let
$w \in \{\text{sess}, \text{hist}\}$ and let
$\widetilde{\mathbf{X}}_u^{\text{ctx},w}$ denote the embedded prefix
within $\widetilde{\mathbf{X}}_u^{w}$: same event-token construction as
the full window, but truncated to timesteps strictly before the target
event. The \emph{target event} is a single (\emph{event type},
\emph{event value}) pair; we treat it as a length-one sequence, embed it
with the same text embedding model, and obtain a one-row matrix
$\widetilde{\mathbf{x}}_{j}$ for candidate $j$. Both the context prefix
and each candidate event follow the same two-stage path: text
$\rightarrow$ pretrained Qwen3-0.6B-embedding
(Section~\ref{sec:method-llm-input}) $\rightarrow$ the corresponding
copy of $g_{\text{seq}}^{\text{LLM}}$ (last-token pooling and projection
as in Eq.~\ref{eq:llm-encoder}). For a user--side pair and a chosen cut
point, this yields a context embedding
$\mathbf{h}_i =
g_{\text{seq}}^{\text{LLM}}(\widetilde{\mathbf{X}}_u^{\text{ctx},w})$
(with the session or history weights matching $w$) and, for each candidate
event $j$, an embedding
$\mathbf{e}_{j}^{\text{evt}} = g_{\text{seq}}^{\text{LLM}}(\widetilde{\mathbf{x}}_{j})$
under the same weights. We index training examples by
$i \in \{1,\ldots,N\}$; each $i$ is one such prefix--target draw from
either $\widetilde{\mathbf{X}}_u^{\text{sess}}$ or
$\widetilde{\mathbf{X}}_u^{\text{hist}}$. The positive label is the true
next event $j_i^+$; we add negatively sampled events to a candidate set
$\mathcal{C}_i$ and set $y_{i,j_i^+}=1$ and $y_{i,j}=0$ for
$j \neq j_i^+$. Rather than a softmax over the full event vocabulary, we
score each candidate with the inner product
$\ell_{i,j} = \mathbf{h}_i^\top \mathbf{e}_{j}^{\text{evt}}$, map it to a
probability with the sigmoid $\hat{p}_{i,j} = \sigma(\ell_{i,j})$, and
optimize binary cross-entropy (Eq.~\ref{eq:llm-bce}), which keeps the
head small and robust as the event space grows. Last-token pooling on the
context prefix encourages $\mathbf{h}_i$ to summarize intent within the
chosen window rather than only local patterns. For content retrieval, we load the session and
history encoders into the user tower and fine-tune them jointly with the
two-tower recommendation objective.
\begin{equation}
  \mathcal{L}_{\text{seq}}^{\text{LLM}} = -\frac{1}{N}\sum_{i=1}^{N}
  \sum_{j \in \mathcal{C}_i}
  \Big[
    y_{i,j}\log \hat{p}_{i,j}
    + (1-y_{i,j})\log(1-\hat{p}_{i,j})
  \Big],
  \label{eq:llm-bce}
\end{equation}
At retrieval fine-tuning time, encoding the \emph{full}
$\widetilde{\mathbf{X}}_u^{\text{sess}}$ and
$\widetilde{\mathbf{X}}_u^{\text{hist}}$ windows (without truncation)
yields $\mathbf{e}_u^{\text{LLM,sess}}$ and
$\mathbf{e}_u^{\text{LLM,hist}}$ in Eq.~\ref{eq:llm-encoder}.

\subsubsection{Producing Session and History Embeddings}
\label{sec:method-llm-encoder}
Each $g_{\text{seq}}^{\text{LLM}}$ consumes the embedded event
sequences $\widetilde{\mathbf{X}}_u^{\text{sess}}$ and
$\widetilde{\mathbf{X}}_u^{\text{hist}}$ from
Section~\ref{sec:method-llm-input} (not the symbolic event-token lists
$\widetilde{S}_u$ and $\widetilde{H}_u$ alone):
the linear input map and adapted backbone above, last-token pooling on
the normalized final hidden state, and a linear projection to the
dimension required by $f_{\text{user}}$. We build two copies---session
and history---both initialized from the same pretrained checkpoint but
fine-tuned separately; after training they share architecture but not
weights. The resulting vectors are
\begin{equation}
  \mathbf{e}_u^{\text{LLM,sess}} =
  g_{\text{seq}}^{\text{LLM}}(\widetilde{\mathbf{X}}_u^{\text{sess}}),
  \qquad
  \mathbf{e}_u^{\text{LLM,hist}} =
  g_{\text{seq}}^{\text{LLM}}(\widetilde{\mathbf{X}}_u^{\text{hist}}),
  \label{eq:llm-encoder}
\end{equation}
where the session and history encoders share the notation
$g_{\text{seq}}^{\text{LLM}}$ but use separate fine-tuned weights, and
$\mathbf{e}_u^{\text{LLM}} = [\mathbf{e}_u^{\text{LLM,sess}};
\mathbf{e}_u^{\text{LLM,hist}}]$ is concatenated into $f_{\text{user}}$
(Eq.~\ref{eq:user-tower}).

\section{Online Serving}
\label{sec:serving}

This section describes how \emph{user-side} representations in
Eqs.~\ref{eq:seq-encoder}--\ref{eq:user-tower}---including
$g_{\text{seq}}$ outputs and LLM-derived auxiliary vectors
(Section~\ref{sec:method-encoder})---are kept fresh and supplied to the
two-tower retrieval model without running heavy encoders on the
user-facing critical path.

\subsection{Nearline Architecture}

A daily batch refresh of user embeddings would be too stale, since
in-session intent can shift within minutes; running $g_{\text{seq}}$,
the LLM encoder, and related sequence pipelines synchronously on each
request is also infeasible given our latency budget ($\text{P95} \le
100\,\text{ms}$ / $\text{P99} \le 200\,\text{ms}$, with only
$\le 50\,\text{ms}$ allotted to model inference). We therefore adopt a
\emph{nearline} design: user-side encoders run asynchronously, shortly
after new activity is observed, and write
$\mathbf{e}_u^{\text{sess}}$, $\mathbf{e}_u^{\text{hist}}$, and
$\mathbf{e}_u^{\text{LLM}}$ (when enabled) to a low-latency store that
the retrieval path reads from synchronously, decoupling heavy encoding
from the latency-sensitive scoring step.

\paragraph{Amazon SageMaker serving.}
We build this on Amazon SageMaker as the model-serving substrate, running
inference on AWS G7 instances equipped with NVIDIA RTX PRO 4500 Blackwell
GPUs, using two deployment modes:
\begin{itemize}
  \item \textbf{Real-time inference.} The synchronous retrieval
  endpoint (Section~\ref{sec:method-retrieval}) is an auto-scaling
  SageMaker deployment. It takes a JSON request with request-time
  non-sequential user features (Section~\ref{sec:method-data}) and
  the latest nearline session, history, and LLM vectors from the
  embedding store, runs $f_{\text{user}}$ to form $\mathbf{e}_u$
  (Eq.~\ref{eq:user-tower}), and returns Top-$K$ items by
  $\mathrm{score}(u,i)$ via approximate nearest-neighbor search over
  a precomputed index of item embeddings from $f_{\text{item}}$
  (Eq.~\ref{eq:item-tower}), without re-encoding the catalog on each
  request.
  \item \textbf{Asynchronous / nearline inference.} User-side encoders
  ($g_{\text{seq}}$, the LLM path, and related jobs) are deployed off
  the critical path, triggered by new activity rather than by a
  recommendation request. This lets us scale retrieval latency and
  embedding refresh throughput independently.
\end{itemize}

\begin{figure*}[t]
  \centering
  \includegraphics[width=0.6\textwidth]{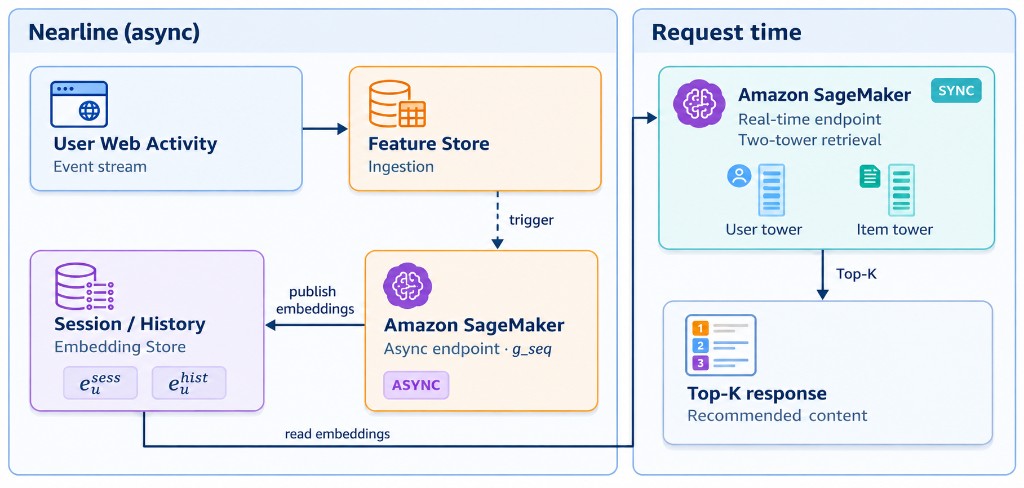}
  \caption{Nearline computation pipeline. New activity triggers
  asynchronous SageMaker endpoints that refresh nearline user-side
  embeddings (including $g_{\text{seq}}$ and LLM outputs) in an
  embedding store; a separate real-time endpoint reads those vectors,
  fuses them in $f_{\text{user}}$, and scores Top-$K$ candidates via
  ANN without heavy encoders on the critical path.}
  \Description{Flowchart of nearline user-embedding jobs triggered by
  new activity, writing vectors to a store, and a real-time endpoint
  that reads those vectors and returns top-ranked recommendations.}
  \label{fig:serving}
\end{figure*}

\paragraph{Nearline computation pipeline.}
Figure~\ref{fig:serving} shows the end-to-end flow. New User Web
Activity events (Section~\ref{sec:method-data}) are ingested into our
feature store (with separate session and history feature groups), which
triggers asynchronous
user-embedding endpoints on refreshed $S_u$ and $H_u$ (including
$g_{\text{seq}}$ and the LLM encoder). Those jobs write
$\mathbf{e}_u^{\text{sess}}$, $\mathbf{e}_u^{\text{hist}}$, and
$\mathbf{e}_u^{\text{LLM}}$ (when enabled) to an embedding store keyed
by user id---the hand-off between nearline encoding and real-time
retrieval.

\subsection{Serving Contract, Freshness, and Reliability}

\paragraph{Serving contract with the application server.}
At request time, the application server does not invoke
$g_{\text{seq}}$ or the LLM encoder: it reads the latest nearline
user-side vectors from the embedding store (key-value lookup by user
id), attaches request-time non-sequential user features
(Section~\ref{sec:method-data}), and forwards a single JSON payload
to the real-time retrieval endpoint for fusion and ANN scoring. Sync-path
latency is bounded by $f_{\text{user}}$, ANN lookup, and ranking overhead,
not by re-encoding full behavioral sequences. The same nearline design
can be extended to the entire $f_{\text{user}}$. Through iteration, we
have since moved to this architecture, as we found that it further
lowers the online latency while maintaining online model accuracy.

\paragraph{Freshness, fallback, and reliability.}
Because updates are asynchronous, there is a bounded staleness window
between a user action and its effect on nearline user embeddings.
If a fresh vector is not yet available (e.g.\ a new session, or a
pipeline delay), we fall back to the most recent cached session/history
(and LLM, if enabled) vectors, or omit optional LLM features, rather
than failing the request. This freshness-versus-cost trade-off is a deliberate design
knob, and we revisit its implications for larger-scale platforms in
Section~\ref{sec:limitations}.

\section{Experiments}
\label{sec:experiments}

We validate LSER with offline retrieval evaluation on logged
interactions (Section~\ref{sec:experiments-offline}) and with a
production online A/B test (Section~\ref{sec:experiments-online}).
Offline experiments isolate auxiliary LLM-derived user features under
controlled settings across the full content catalog; online experiments measure
engagement and product-level conversion on live traffic.

\subsection{Offline Evaluation}
\label{sec:experiments-offline}

We assess retrieval quality offline before and alongside production
experiments. All models are trained and evaluated on logged
(user, item) interaction data from autodesk.com.

\paragraph{Data collection and splits.}
We assemble supervised examples from a recent, fixed calendar window of
logged autodesk.com activity, built through the label-receipt pipeline
(Section~\ref{sec:method-labels}) together with the user-side features
in Section~\ref{sec:method-data}. The exact window length is held
constant across all offline runs compared in this paper. Each row is one label receipt
and thus one supervised $(u, i^+)$ pair; if a user generates several
qualifying engagement events (including multiple events within the same
browsing session), each event that yields a receipt becomes a separate
training or evaluation row with its own timestamp and feature snapshot.
Samples are ordered by event time; we apply a \emph{temporal} train/validation/test
partition so that models are never trained on future events. Concretely,
after sorting by timestamp, the earliest 80\% of samples form the
training set, the next 10\% validation, and the final 10\% test---a
standard chronological 80/10/10 split. Hyperparameter tuning and early
stopping use validation; all retrieval metrics reported for model
comparison are computed on the held-out test split only. For each
receipt row, session and history sequences are reconstructed
point-in-time at that row's label timestamp using the rules in
Section~\ref{sec:method-sequences} (same point-in-time rules and
feature construction logic as in production).

\paragraph{Training setup.}
All ablation variants use the same two-tower architecture and are
trained jointly end-to-end on the training split with the InfoNCE
objective (Eq.~\ref{eq:infonce}) and in-batch negatives. Each train and
test row is a fused positive $(u, i)$ pair from the same label-receipt
construction as in Section~\ref{sec:method-labels}. Validation loss
and retrieval metrics on the validation split guide checkpoint
selection; we report numbers only on the chronologically held-out test
split. Item embeddings for evaluation are precomputed with the trained
item tower and indexed for full-catalog retrieval, matching the
production retrieve-then-rank pattern in Section~\ref{sec:method-retrieval}.
Training runs on AWS G5 instances equipped with NVIDIA A10G GPUs, and
data loading and preprocessing are parallelized across multiple CPU
workers to reduce GPU idle time.

\paragraph{Retrieval metrics.}
Each test row defines one retrieval evaluation: given the user features
and sequences available at that row's label time, we embed the user with
$f_{\text{user}}$, score every item in the evaluation catalog by cosine
similarity to $\mathbf{e}_u$, and ask whether the row's logged positive
$i^+$ appears in the top-$K$ list. We do \emph{not} aggregate multiple
positives into a single query-level relevant set; instead, multiple
clicks or engagements in the same session are represented as multiple
receipt rows, and metrics are computed by averaging over rows. \textbf{Hit Rate@$K$}
(HR@$K$) is the fraction of test rows for which $i^+$ appears in the
top-$K$ retrieved items; with a single clicked item per row, HR@$K$ is
numerically identical to Recall@$K$. We report HR@$K$ for
$K \in \{1, 5, 10\}$ and \textbf{mean reciprocal rank (MRR)}. For each
test row, MRR uses the rank of $i^+$ in the scored candidate list (one
plus the number of items ranked above the positive); MRR is the mean of
$1/\text{rank}$ over rows. Higher MRR indicates that logged clicks
appear nearer the top of the retrieval list.

\paragraph{Reporting treatment effects.}
Table~\ref{tab:offline-ablation} reports \emph{absolute} changes on
the held-out test split (treatment minus baseline): HR@$K$ in
percentage points (pp) and MRR as a difference on the usual $[0,1]$
scale. Table~\ref{tab:ab-results} reports \emph{relative lifts} of
treatment over control,
$(m_{\mathrm{trt}} - m_{\mathrm{base}}) / m_{\mathrm{base}} \times 100\%$,
for online engagement and conversion metrics.

\paragraph{Retrieval protocol.}
Offline retrieval uses the same scoring and surface-aware candidate
rules as production: each test row is evaluated against the indexed
catalog allowed for the \emph{recommendation surface} tied to that label
(on-page slot / element location, not item content type alone). We
report metrics in aggregate and by surface where sample size supports
stable estimates; Table~\ref{tab:offline-ablation} focuses on popular
products and customer success, which also appear among our labeled
surfaces in offline data.

\paragraph{LLM-derived feature ablation.}
The user tower combines multiple learned sequence encoders with dense
context features. We ablate \emph{auxiliary} LLM-derived session and
history embeddings: two 64-dimensional vectors
($\mathbf{e}_u^{\text{LLM,sess}}$, $\mathbf{e}_u^{\text{LLM,hist}}$)
concatenated into $f_{\text{user}}$ (Eq.~\ref{eq:user-tower}) while
holding $g_{\text{seq}}$ and all other features fixed. The
\textbf{baseline} omits $\mathbf{e}_u^{\text{LLM}}$; the
\textbf{treatment} appends it. Both variants use the same label-receipt
data, chronological split, InfoNCE training, item index, and
offline protocol above; only the LLM feature toggle differs.

Table~\ref{tab:offline-ablation} highlights surfaces where the LLM
toggle improves offline HR@$K$ and MRR. \emph{Remaining} surfaces and
metrics showed no material
improvement---retrieval quality on those surfaces remained essentially
flat versus the baseline.

\begin{table}[t]
  \caption{Offline ablation: adding LLM-derived 64-d session/history
  vectors as auxiliary user features on the held-out test split.
  Absolute changes (treatment $-$ baseline): $\Delta$HR@$K$ in
  percentage points; $\Delta$MRR on $[0,1]$. Other surfaces and metrics
  are not tabulated because they remained essentially flat relative to
  baseline.}
  \label{tab:offline-ablation}
  \small
  \begin{tabular}{lcc}
    \toprule
    Metric & Popular products & Customer success \\
    \midrule
    $\Delta$HR@1  & $+0.2\%$ & $+6.0\%$ \\
    $\Delta$HR@5  & $+2.2\%$ & $+2.4\%$ \\
    $\Delta$HR@10 & $+9.6\%$ & --- \\
    $\Delta$MRR   & $+0.021$ & $+0.030$ \\
    \bottomrule
  \end{tabular}
\end{table}

The largest lift is on popular products, where LLM features raise
HR@10 by $+9.6$~pp and MRR by $+0.021$, with non-zero HR@1/5 where the
baseline often misses entirely. Gains on customer success stories are
concentrated in top ranks (HR@1 $+6.0$~pp, MRR $+0.030$). These two surfaces also relate to modules measured in our
online A/B test (Section~\ref{sec:experiments-online}). On other
element locations and metric pairs, offline scores stayed essentially
flat versus baseline. The
ablation therefore supports adding LLM-derived
signals where catalog breadth and sparsity stress the learned encoders
alone, without claiming uniform lift on every surface or content type.

Based on how the two paths are built, we speculate that the lift comes
from two sources. First, the LLM path sees richer inputs: $g_{\text{seq}}$
consumes only events that resolve to a catalog item id (others are
dropped or mapped to an unknown token), whereas the LLM encoder
serializes the (\emph{event type}, \emph{event value}) text of logged
events across a broader range of content and page types, including pages
with no catalog item id (Section~\ref{sec:method-llm-input}). Second,
text embeddings carry semantic signal that can generalize to sparse or
long-tail content whose id-based representations are weakly trained,
which is consistent with the lift appearing on surfaces with a broad,
sparse catalog.

\subsection{Online A/B Testing}
\label{sec:experiments-online}

We evaluate production impact with a randomized online A/B test on
autodesk.com. Eligible traffic was split evenly (50\%/50\%) between
control and treatment for the experiment window; we log clicks on
recommended content and downstream conversion events on the surfaces
where recommendations are rendered.
Table~\ref{tab:ab-results} summarizes \emph{relative lifts} of the
treatment over control for three live surfaces: the homepage product
carousel, the homepage industry carousel, and AutoCAD customer stories
on the ACD Product Center page. Reported metrics include
click-through rate on recommended tiles (product/industry combined and
customer stories separately), and product-level trial start and install
rates and product order submission rate (downstream of product
engagement).

The A/B test compares the \emph{control} arm (prior rule-based content
engine: fixed business rules, no learned session representation) with
the \emph{treatment} arm (initial LSER deployment using the same
two-tower retrieval and nearline user-embedding serving design in
Sections~\ref{sec:method-retrieval} and~\ref{sec:serving}). The
treatment at experiment time used a BERT4Rec-style
$g_{\text{seq}}$~\cite{sun2019bert4rec} in the user tower and did
\emph{not} enable auxiliary LLM-derived features
(Eq.~\ref{eq:user-tower}, Section~\ref{sec:method-encoder}). That
scope reflects which launch experiment we prioritized---validating
learned, session-aware retrieval and nearline user embeddings against
the legacy rule engine---rather than any inability to serve LLM
features: the asynchronous LLM embedding path is already integrated
into the live stack (Section~\ref{sec:serving}) and can be toggled
independently of the rest of the user tower. A dedicated online test
of $\mathbf{e}_u^{\text{LLM}}$ on production traffic is planned as
follow-up work.

Table~\ref{tab:ab-results} shows that this learned, session-aware
retrieval stack delivers statistically significant lifts over the
rule-based baseline on live traffic---a production reference for the
overall LSER design. Separately, our offline ablation
(Section~\ref{sec:experiments-offline}) isolates the incremental value
of \emph{adding} those LLM-derived vectors on top of the same learned
encoders on a heterogeneous, multi-content-type catalog, including
surfaces and item types not all represented in Table~\ref{tab:ab-results}.
Together, the online experiment establishes that moving from rules to
learned retrieval pays off in production; offline results motivate
turning on LLM auxiliary features in a subsequent live experiment.

\begin{table}[t]
  \caption{Online A/B test under a 50/50 traffic split. Relative lifts
  of treatment over the rule-based control. Engagement rows are
  click-through rate on recommended tiles; conversion rows are
  product-level trial and order rates. All metrics were statistically
  significant ($p < 0.05$).}
  \label{tab:ab-results}
  \begin{tabular}{p{0.58\linewidth}r}
    \toprule
    Metric & Relative Lift \\
    \midrule
    \multicolumn{2}{l}{\textbf{Engagement}} \\
    Homepage Prod\&Ind.\ carousel CTR & +10.3\% \\
    AutoCAD customer story CTR        & +36.1\% \\
    \midrule
    \multicolumn{2}{l}{\textbf{Conversion}} \\
    Product trial start rate          & +17.1\% \\
    Product trial install rate        & +10.5\% \\
    Product order submission rate     & +14.8\% \\
    \bottomrule
  \end{tabular}
\end{table}

\section{Discussion}
\label{sec:discussion}
\label{sec:limitations}

While LSER shows encouraging gains in production and offline
evaluation, we highlight three limitations of the present work.

\textbf{Lack of an online ablation on LLM features.} The online A/B test
in Section~\ref{sec:experiments-online} contrasts the rule-based engine
with an initial LSER deployment that uses learned
$g_{\text{seq}}$ (BERT4Rec-style) but does not yet toggle the auxiliary
LLM-derived vectors in Eq.~\ref{eq:user-tower}. That gap reflects
experiment scheduling at launch, not serving feasibility: the
nearline LLM path (Section~\ref{sec:serving}) is integrated into the
live system. Offline evaluation
(Section~\ref{sec:experiments-offline}) shows gains from \emph{adding}
those vectors while holding $g_{\text{seq}}$ fixed. Table~\ref{tab:ab-results}
validates session-aware learned retrieval over rule-based surfacing;
the remaining gap is an online ablation that enables
$\mathbf{e}_u^{\text{LLM}}$ on live traffic, which we plan to run next.

\textbf{Generalizability to larger-scale platforms.} LSER is validated
on autodesk.com, which serves on the order of millions of requests per
day. It remains an open question whether the same retrieval
architecture, session encoder design, and nearline serving approach
would hold up at the scale of large social media or consumer platforms
that operate several orders of magnitude more traffic and a
substantially larger and faster-changing item catalog. In particular,
the nearline serving path, whose freshness/cost trade-off is discussed
above, may need to be redesigned at such scale, since the compute and
infrastructure cost of recomputing session embeddings nearline grows
with traffic volume. We view LSER as evidence that this approach is
viable at our scale, rather than as a claim that it generalizes
unmodified to substantially larger platforms.

\textbf{Limited exploration of the LLM backbone and training
objective.} Our current LLM embedding path uses a single
next-event-prediction objective to fine-tune the session encoder
$g_{\text{seq}}^{\text{LLM}}$ (initialized from a pretrained Qwen3-style
checkpoint). We have not yet explored alternative training objectives
(e.g., contrastive or multi-task formulations) or alternative base
model architectures and sizes, any of which could further improve
embedding quality or efficiency. Systematically exploring this design
space is left to future work.

\section{Conclusion}
\label{sec:conclusion}

We presented LSER, a production retrieval stack for autodesk.com that
adds LLM-derived session and history vectors as auxiliary features in
a two-tower model, with nearline refresh of the full user embedding
stack (learned sequential encoders plus LLM features). We described the
retrieval architecture, feature construction, InfoNCE training
objective, nearline serving on Amazon SageMaker, and an offline retrieval
evaluation (Hit Rate@$K$, MRR). Offline ablation toggles LLM features
across the full multi-content-type catalog; an online A/B test shows
statistically significant gains over a rule-based baseline on live
engagement and product conversion metrics. Together, these results
support augmenting learned retrieval with LLM-derived signals as a
practical path for heterogeneous content recommendation at our scale,
with online LLM-feature ablations and broader backbone exploration as
natural next steps.

\begin{acks}
This work was made possible through a cross-functional effort. We are
deeply grateful to Vishal Rajendran, Audrey Webb, Yathaarth Bhansali,
Sahan Somawardana, Deepthi Viswanathan, Pavel Kulikou, Eduardo
Alburqueque, Rishabh Arora, Ambir Patel, and Stuart Xu, as well as everyone
else who contributed to the design, development, and production
deployment of this system.
\end{acks}

\bibliographystyle{ACM-Reference-Format}
\bibliography{references}

\end{document}